\documentclass[aps,prb,twocolumn,groupedaddress]{revtex4}

\usepackage[pdftex]{graphicx}

\begin{document}
\title{Direct Imaging of Coherent Quantum Transport in Graphene Heterojunctions}
\author{E. D. Herbschleb$^1$, R. K. Puddy$^1$, P. Marconcini$^2$, J. P. Griffiths$^1$, G. A. C. Jones$^1$, M. Macucci$^2$, C. G. Smith$^1$, M. R. Connolly$^{1,3}$}
\affiliation{$^1$Cavendish Laboratory, Department of Physics, University of Cambridge, Cambridge, CB3 0HE, UK\\
$^2$Dipartimento di Ingegneria dell'Informazione, Universit\`a di Pisa, Via G. Caruso 16, I-56122 Pisa, Italy
\\
$^3$National Physical Laboratory, Hampton Road, Teddington TW11 0LW, UK}

\begin{abstract}
 
We fabricate a graphene \textit{p-n-p} heterojunction and exploit the coherence of weakly-confined Dirac quasiparticles to resolve the underlying scattering potential using low temperature scanning gate microscopy. The tip-induced perturbation to the heterojunction modifies the condition for resonant scattering, enabling us to detect localized Fabry-P\'{e}rot cavities from the focal point of halos in scanning gate images. In addition to halos over the bulk we also observe ones spatially registered to the physical edge of the graphene. Guided by quantum transport simulations we attribute these to modified resonant scattering at the edges within elongated cavities that form due to focusing of the electrostatic field. 

\end{abstract}

\maketitle

Developing methods to characterize and preserve the coherence of interacting quantum systems is essential for exploring fundamental problems in quantum mechanics and for realizing novel technologies which operate using entanglement and superposition. In quantum devices based on semiconducting two-dimensional electron gases, spatial coherence of the electron field can be visualized directly by scanning a sharp metallic tip over the surface while measuring its conductance, in a technique known as scanning gate microscopy (SGM). SGM images of interfering electron waves provide exquisite real-space information that can be used to diagnose scattering and decoherence mechanisms stemming from the underlying material \cite{Crook2000, Topinka2001, Aidala2007, Crook2003a}. In the new breed of quasi-two-dimensional Dirac materials such as graphene and the surface states of topological insulators, where low-energy quasiparticle excitations mimic a two-dimensional gas of relativistic chiral charged neutrinos \cite{Novoselov2005}, signatures of coherence in bulk transport measurements have unravelled the complex interplay between this bandstructure and elastic scattering rates \cite{Tikhonenko2009}. However, while SGM has been used to image mesoscopic doping inhomogeneities \cite{Connolly2010, Jalilian2011}, edge effects \cite{Chae2012}, localized states \cite{Connolly2011, Schnez2010}, and quantum interference \cite{Berezovsky2010a, Berezovsky2010c}, coherent scattering within a tailored scattering potential has not been characterized or exploited using local probes. Graphene devices are now ripe for such probing techniques, especially with the opportunity to image novel physical effects such as Vesalago lensing \cite{Cheianov2007}, cloaking \cite{Gu2011a}, and superlattice collimation \cite{Park2008}.


In this letter we demonstrate how the spatial coherence of Dirac quasiparticles within a \textit{p-n-p} heterojunction can be exploited to resolve the scattering potential in a graphene monolayer by SGM. Our solution is paradigmatically similar to experiments where narrow nanofabricated gates enable the effects of coherence and Klein tunnelling to be explored, even in low-mobility samples \cite{Young2009}. Due to interference between electron waves scattered from its boundaries, the conductance of a heterojunction exhibits periodic oscillations as a function of the local Fermi wavelength \cite{Shytov2008, Huard2007, Gorbachev2008, Nam2011, Young2009}. While mapping these resonances by SGM would provide information about the potential landscape, the presence of metallic top gates has so far prohibited this. Here we employ an \textit{in-situ} electrostatic patterning technique \cite{Connolly2012, Crook2003} to fabricate the heterojunction and spatially resolve cavities in the scattering potential through the presence of halos - spatially distinct ring structures - where the resistance of the heterojunction is higher than the background. In addition to identifying a sequence of Fabry-P\'{e}rot (FP) halos that stem from multiple disorder-induced cavities in the bulk of the well, we image narrow and highly periodic resonances that are registered to the physical edges of the graphene flake. We attribute these resonances to the enhanced electrostatic coupling at the edges, that results from focusing of the electric field \cite{Silvestrov2008}.

\begin{figure}[!t]
\includegraphics[width=85mm]{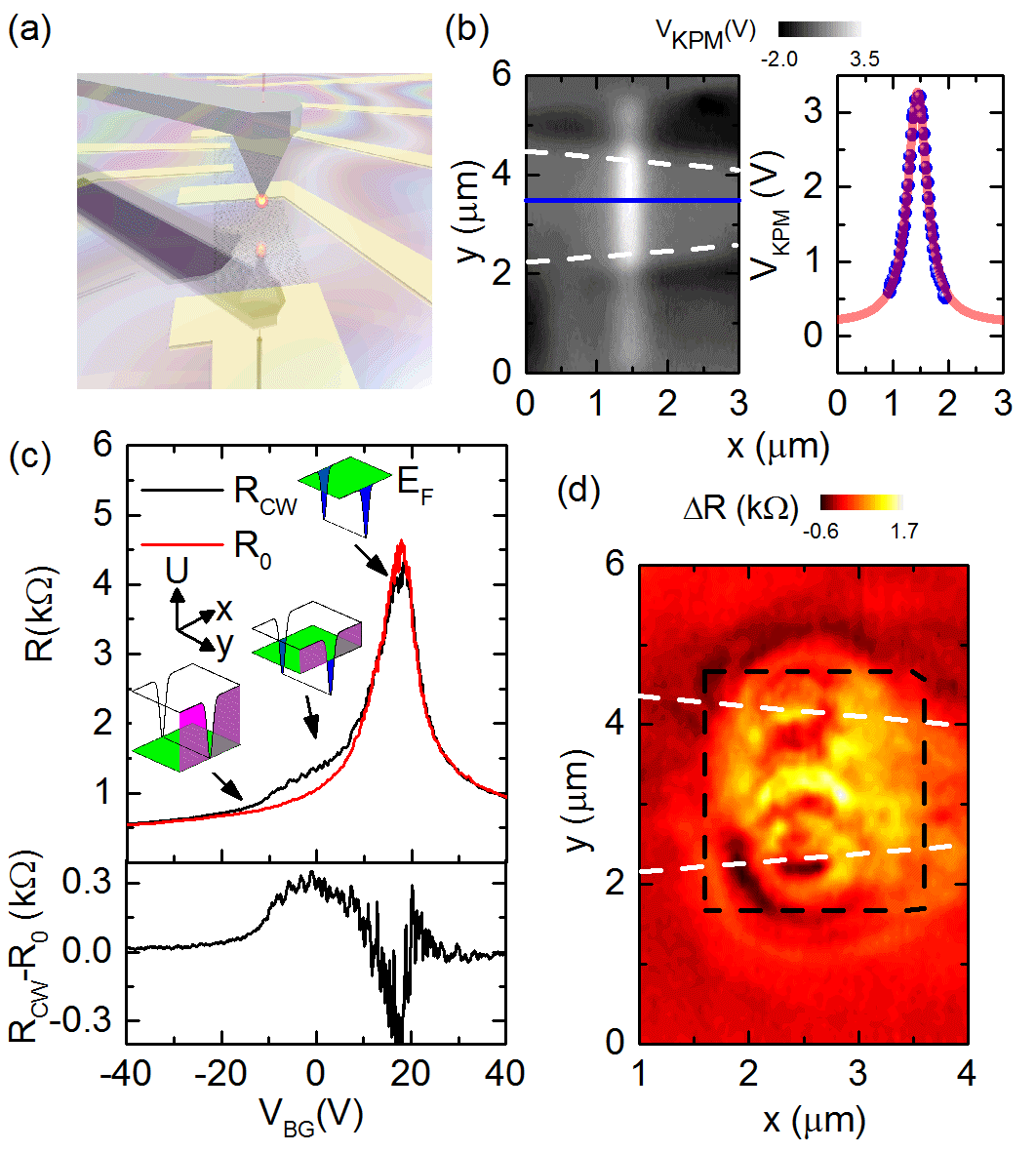}
\caption{(a) Schematic setup of charge writing on a graphene device. (b) Kelvin probe image (left) and line profile (right - blue points: experimental data, red line: Lorentzian fit) captured at a lift height of 100 nm after charge writing. White dashed lines indicate the edge of the flake. (c) Three terminal resistance as a function of back-gate voltage before (R$_0$) and after (R$_{CW}$) charge writing. Cartoon insets illustrate the relation between the schematic potential landscape after charge writing, and the Fermi level at different back-gate voltages. Lower panel shows the change in resistance. (d) Scanned gate micrograph over the region shown in (b). The black dashed outline indicates the region used for higher resolution scans. All measurements at $T\approx$4.2 K.}    
\label{Fig:Fig1}
\end{figure}

Fig. \ref{Fig:Fig1}(a) shows a schematic of the configuration used for our experiments \cite{Connolly2012}. We fabricate a graphene heterojunction at low temperature ($T$$\approx$4.2 K) by depositing a line of charge into a dielectric coating over the graphene using triboelectrification (See Supplementary Material and Ref. \cite{Zhou2013}). An image of the surface potential measured using Kelvin probe microscopy (KPM) is shown in Fig. \ref{Fig:Fig1}(b). The KPM line profile is well fitted, at $V_{BG}$=0 V, by a Lorentzian with full-width half maximum of $\approx$200 nm. The effect of the written charge is similar to a conventional top gate and is revealed in transport by an increase in the resistance at around $V_{BG}\approx$0 V, the local neutrality point under the charge, and a decrease around the original neutrality point [Fig. \ref{Fig:Fig1}(c)]. The mean change in resistance $\Delta R(V_{BG})=R_{CW}-R_0$ is plotted in the lower panel and is well predicted by a simple classical model for diffusive transport (see Supplementary Material). 

\begin{figure*}
\includegraphics[width=180mm]{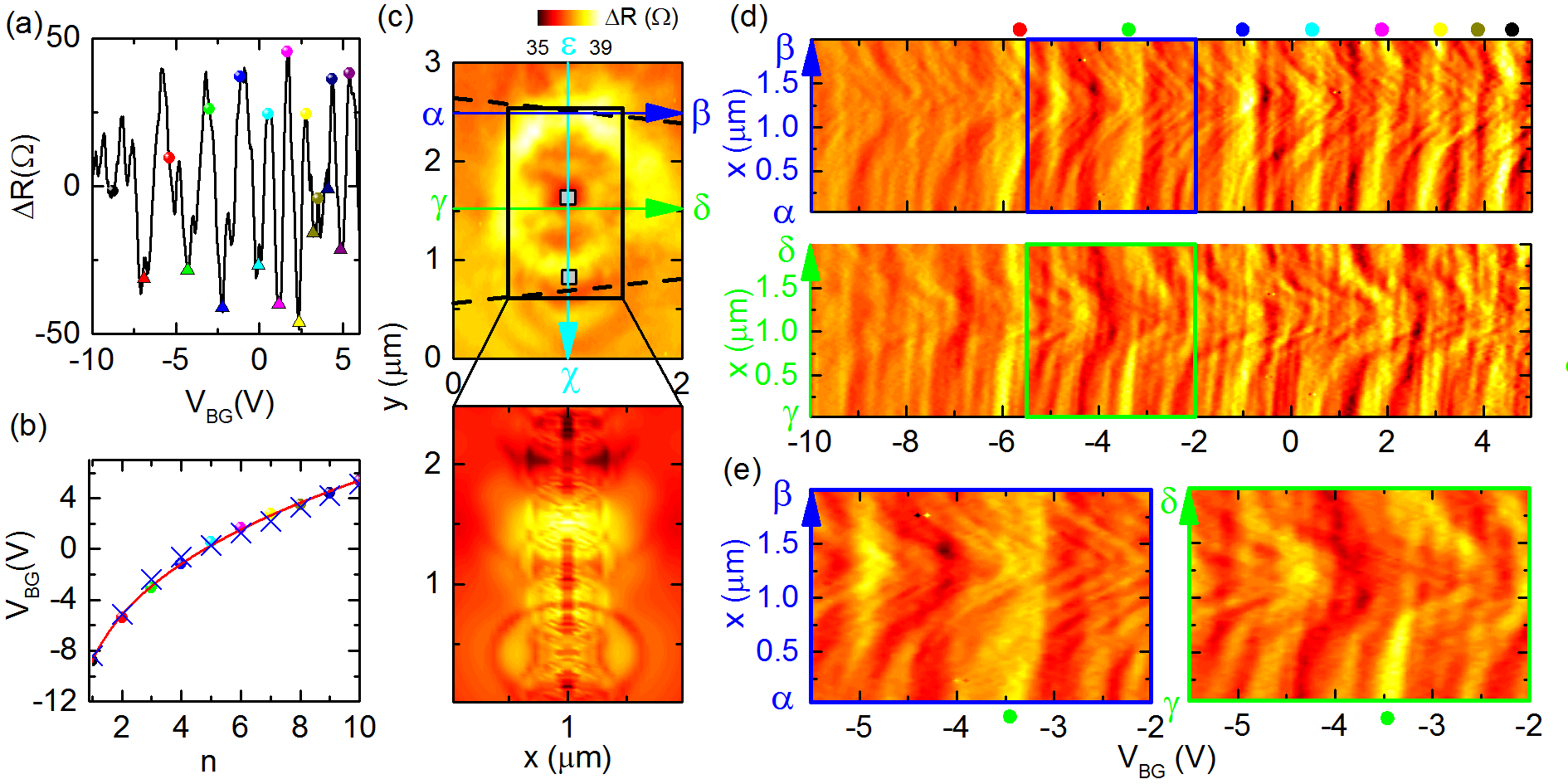}
\caption{(a) Change in resistance measured as a function of back-gate voltage and detrended in order to emphasise the peaks (circles) and troughs (triangles). (b) Back-gate voltage of the peak positions in (a) as a function of index. Solid line is a best fit to the data and crosses are the result of the quantum transport simulations. (c) Measured scanning gate image at $V_{BG}=0 V$ (upper)and a typical simulated image within the same regime (lower). The tip voltage is -10 V in both cases. Blacked dashed lines indicate the edge of the flake. (d) Derivative of the detrended resistance with respect to the back-gate voltage, plotted as a function of back-gate voltage and tip position along $\alpha \rightarrow \beta$ and $\gamma \rightarrow \delta$. Coloured circles correspond to the peaks indicated in (a). (e) Higher resolution of the range highlighted by boxes in (d). All measurements at $T\approx$4.2 K.}    
\label{Fig:Fig2}
\end{figure*}

Superimposed on this broad background modulation of the resistance are reproducible oscillations that develop for back-gate voltages greater than $\approx$-10 V. To reveal their microscopic origin we fix the back-gate deep within the \textit{p-n-p} regime ($V_{BG}$=0 V) and monitor the two-terminal resistance while the static tip is scanned at a lift height of $\approx$130 nm over the dielectric. A typical scanned gate image [Fig. \ref{Fig:Fig1}(d)] shows a nest of circular features with different focal points centred over the heterostructure with little contrast outside this region. Such circular halos are frequently observed in SGM images and their spatial registration to a specific area within a nanodevice is typically attributed to tip-induced resonant tunnelling of individual charges through quantum dots \cite{Woodside2002}, to interference of electron waves at that point \cite{Martins2013}, or to Fabry-P\'{e}rot resonance between the tip and a scattering potential \cite{Topinka2001}. By identifying a correlation between the resistance oscillations in Fig. \ref{Fig:Fig1}(c) and the halos in Fig. \ref{Fig:Fig1}(d), we demonstrate they are entirely consistent with tip-induced perturbation to quantum interference within the heterojunction itself.

To emphasise the resistance oscillations we detrend the data by subtracting a smoothed resistance and obtain the $\Delta R(V_{BG})$ shown in Fig. \ref{Fig:Fig2}(a). Despite the obvious disorder, a clear sequence of roughly 10 resonances emerge, with amplitudes of $\approx$50 $\Omega$ and periodicity $\Delta V_{BG}\approx$1-4 V over the range of back-gate voltage from -10 V to 5 V, beyond which they become indistinguishable from the aperiodic conductance fluctuations. We estimate the back-gate voltage of the resistance maxima by smoothing the raw data and identifying a regular sequence of dominant peaks and troughs indicated by the circles and triangles in Fig. \ref{Fig:Fig2}(a). The peak separation is roughly linear at high energy [Fig. \ref{Fig:Fig2}(b)]. Such resonances are consistent with previous studies and point to interference effects between electron waves scattered at the \textit{p-n} junctions that define the heterojunction $\cite{Young2009}$. We can estimate the expected period in the linear regime by assuming the \textit{p-n} interfaces are separated by a distance $L$ and that the phase accumulated by an electron traversing the cavity is $\phi=2k_xL$. In graphene this leads to the relationship $\Delta n = 2 \sqrt{\pi n}/L$, where $n$ is the carrier density within the heterojunction and is conventionally controlled by a top gate \cite{Young2009}. Note that there is some uncertainty in the local Dirac point of the cavity and our data is likely to depart from this $\Delta n \propto \sqrt{n}$ dependence because the back gate also modifies the global carrier density. Nonetheless, based on the assumption the local Dirac point is between $V_{BG}=$-10 and -5 V, and at $V_{BG}$=4 V we have $n\approx$0.5 - 1$\times$10$^{12}$ cm$^{-2}$, we derive a periodicity of the FP oscillations of $\Delta V_{BG}$=2 V for a cavity $L$$\approx$200 nm. While this estimate is in broad agreement with the experimental value for the peak separation, the origin of the smaller fluctuations and the non-uniform fringe visibility betrays the influence of disorder. 	

\begin{figure}[!t]
\includegraphics[width=85mm]{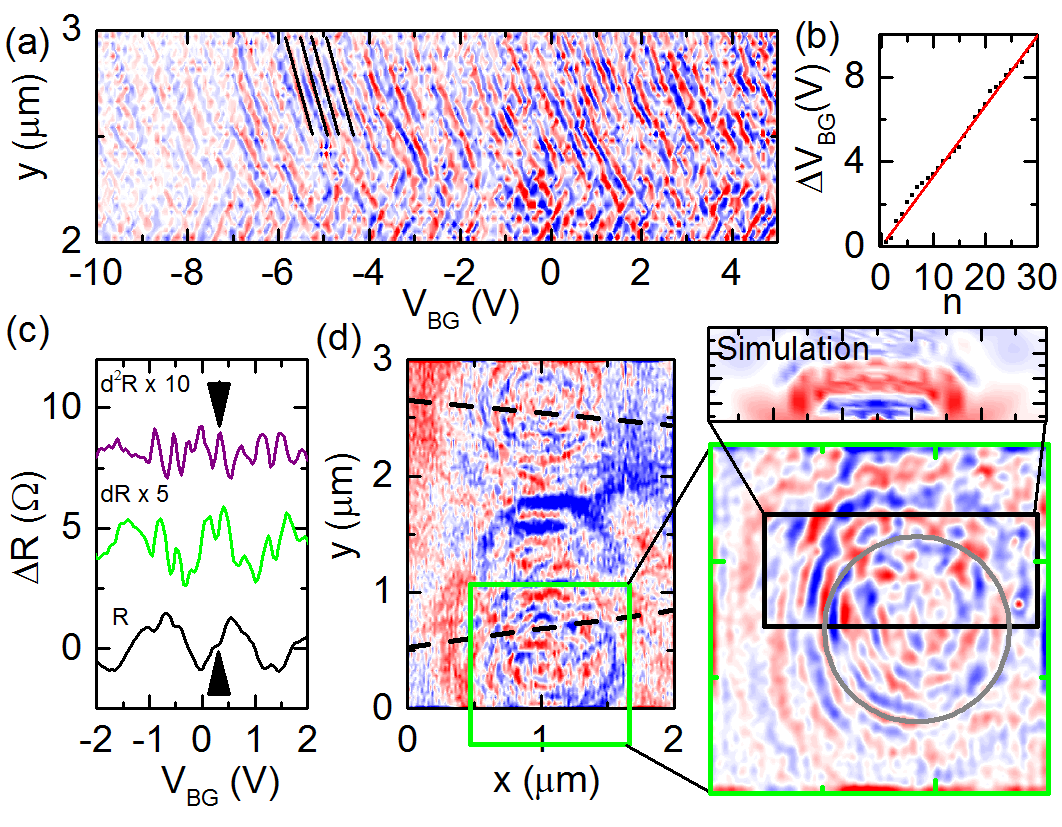}
\caption{(a) Second numerical derivative ($d^2 R$) of the resistance as a function of back-gate voltage and tip position along the direction $\varepsilon \rightarrow \chi$. Solid black lines indicate the uniform shift caused by moving the tip closer to the edge. (b) Plot showing the back-gate voltage of each peak in $d^2 R$. The solid line is the best fit to the data. (c) Plot showing the relationship between the raw back-gate trace and the numerical derivative. Peaks in $d^2 R$ correspond to dips in the raw data. (d) Difference image constructed by subtracting two raw SGM images captured at two different back-gate voltages. In the right panel we show a higher-resolution image for the dashed box shown in the left panel, along with a simulated SGM image showing a halo over the edge. All measurements at $T\approx$4.2 K.}    
\label{Fig:Fig3}
\end{figure}	

In order to correlate the halos in Fig. \ref{Fig:Fig2}(c) with the putative FP resonances identified in Fig. \ref{Fig:Fig2}(a), we choose y-positions over the edge ($\alpha$) and middle ($\gamma$) of the flake, and sweep the back-gate voltage with the tip parked at $x$-positions along the lines $\alpha \rightarrow \beta$ and $\gamma \rightarrow \delta$ in Fig. \ref{Fig:Fig2}(c). The resulting data is shown in Fig. \ref{Fig:Fig2}(d), where the numerical derivative $d \Delta R/dV_{BG}$ has been plotted to add emphasis to the location of the peaks. In both cases, as the tip approaches the heterojunction, a subset of the resonances that are visible even in the absence of the tip undergo a shift in back-gate voltage, while others are less affected. Since only resonances which shift give rise to halos, from these data we deduce that the FP resonances identified in Fig. \ref{Fig:Fig2}(a) correlate with halos centred over the middle of the flake ($\gamma \delta$). We provide evidence for this in the case of a particular resonance marked by a green circle ($V_{BG}\approx$-3.5 V) in Fig. \ref{Fig:Fig2}(e), which shows a narrower range of $V_{BG}$. Along the line $\alpha \beta$ this resonance is only weakly affected by the tip, while along $\gamma \delta$ it is fully perturbed and gives rise to halos similar to the one visible in Fig. \ref{Fig:Fig2}(c). While this pattern can be confirmed by inspection for the majority of peaks along $\alpha \beta$ in Fig. \ref{Fig:Fig2}(d), a number of resonances are unpertubed which suggests they derive from a different location within the heterojunction. Indeed, in Fig. \ref{Fig:Fig2}(c) we can clearly identify 2 focal points separated by roughly 1 $\mu$m [grey squares, Fig. \ref{Fig:Fig2}(c)]. From this we infer that the FP resonances arise from cavities distributed along the heterojunction. While the entrance and exit barriers are sufficiently correlated to define cavities with a well-defined length, each resonates at a slightly different back-gate voltage owing to the different doping and local Fermi wavelength, leading to a broadening and amplitude-suppression of the peaks \cite{Shytov2008}.  

This finding is fully supported by our numerical quantum transport simulations (see Supplementary Material). Due to the relatively large size of the graphene flake under investigation, we employ a code \cite{Logoteta2014} based on a continuum, envelope function formulation \cite{Marconcini2011}, instead of an atomistic approach. Furthermore, in order to easily examine many possible potential landscapes and the outcome of scanning probe experiments, we adopt, under the hypothesis of a slow-varying potential, a simplified procedure for the approximate evaluation of the potential profile as a function of the bias voltages \cite{Marconcini2014}. In our simulations we are able to identify resonances when the Lorentzian-shaped cavity is partitioned into several cavities in parallel by narrow longitudinal walls with a height corresponding to about $10$\% of the cavity depth. The numerical results for the peak spacing is shown alongside the experimental results in Fig. \ref{Fig:Fig2}(b). In the lower panel of Fig. \ref{Fig:Fig2}(c) we present the resistance computed as a function of the position of a scanning probe tip, which is scanned over the flake. The result is rather close to the experimental data and supports our interpretation that multiple coherent subcavities are present in the experimental situation. We note that despite the diffusive nature of transport and the relatively high temperature, our findings are consistent with phase coherent effects as both the thermal length $L_{T} = \pi^{1/2}\hbar^2k_F/4	m^*k_B T \approx$ 1.1 $\mu$m, where $k_F$ is the Fermi wavevector and $m^*$ is the graphene effective mass \cite{Shaw2001}, and the dephasing length $L_{\phi} \approx$300 nm, measured via weak localisation, are longer than the cavity.	

\begin{figure}[!b]
\includegraphics[width=85mm]{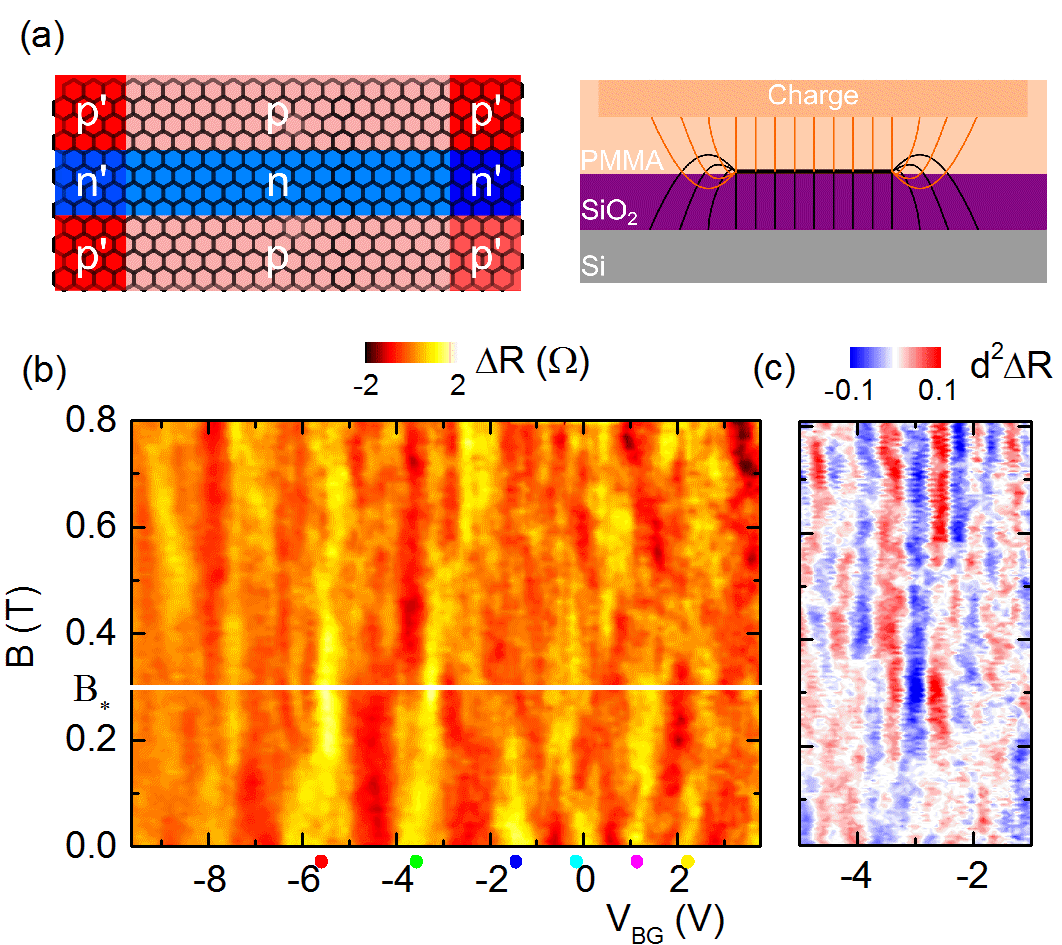}
\caption{(a) Schematic of the inhomogeneous carrier density distribution induced by the electric field focusing at the edge of the flake, such that $p'(n')>p (n)$. For clarity the field lines associated with the charge in the dielectric coating and with the back-gate have been drawn separately. (b) Detrended resistance as a function of back-gate voltage and magnetic field. $B_*$ indicates the field at which the FP resonance are suppressed. (c) Second numerical derivative of the data in (b).}    
\label{Fig:Fig5}
\end{figure}

Closer inspection of data in Fig. \ref{Fig:Fig2}(d) reveals a striking sequence of resonances that are strongly perturbed when the tip is over the edge. To make them more pronounced we stepped the tip parallel to the QW, along the line $\varepsilon \rightarrow \chi$ in Fig. \ref{Fig:Fig2}(c), and display a section captured close to the edge in Fig. \ref{Fig:Fig3}(a). The resonances are made more visible by taking the numerical second derivative ($d^2 R$) of the raw data. The succession of peaks and troughs is highly reproducible, periodic over the full range of back-gate voltage, and shift uniformly as a function of tip position. In order to locate the focal point of the finer resonances in SGM images, we examine difference images constructed by subtracting two images captured at two values of $V_{BG}$. Since the main FP peaks do not change appreciably they are effectively eliminated from the image. Fig. \ref{Fig:Fig3}(d) shows clearly that the finer resonances are centred over the edge of the flake. These resonances have an average period of $\approx$0.33 V in back-gate voltage [Fig. \ref{Fig:Fig3}(b)], which is a factor of three smaller than the main cavity resonances. Another distinguishing feature of the edge resonances is rather their shallower amplitude, which is an an order of magnitude less. Indeed, in the raw data the peaks in $d^2 R$ actually correspond to small dips [Fig. \ref{Fig:Fig3}(c)]. 

Our observations are consistent with the formation of modified FP cavities at the edges of the flake [Fig. \ref{Fig:Fig5}(a)]. For a given cavity length, the back-gate voltage separation between consecutive peaks decreases if the capacitive coupling $\alpha$ increases owing to the accelerated movement of the Fermi energy through the quasi-bound energy levels. It is now well-established that $\alpha$ is enhanced at the edges due to electric field focusing \cite{Silvestrov2008}. The resulting period in back-gate voltage is consequently shorter by a factor $\alpha_{edge}/\alpha_{bulk}$, which our data implies is $\approx$3. This is in excellent agreement with our electrostatic simulations as well as capacitance measurements in the quantum Hall regime \cite{Vera-Marun2013}. We also confirmed that, due to the varied capacitive coupling, the accumulated charge alters the halos associated with the cavities at the edges by performing simulations with and without charge accumulation and inspecting the difference. The result for the lower edge of the flake is shown in Fig. \ref{Fig:Fig3}(d), where halos centered on the accumulation-induced edge cavity clearly appear. 

Another distinction between the bulk and edge resonances is their behaviour in a perpendicular magnetic field, shown in Fig. \ref{Fig:Fig5}(b). Although disorder within the bulk cavities prevents us from observing a clear Berry-phase induced shift of the bulk resonances to lower carrier density \cite{Young2009}, as expected from the curvature of semi-classical quasiparticle trajectories in a magnetic field, the interference condition for FP resonance ceases to be satisfied and the regular sequence of conductance oscillations is destroyed above a field $B_* \approx$ 300 mT [Fig. \ref{Fig:Fig5}(b)]. Within the framework presented in Ref. \cite{Cheianov2006}, $B_*\propto \sqrt{k_F/\pi L^2d}$, where $d$ is the characteristic lengthscale of the \textit{p-n} junction, and $k_F=\sqrt{n \pi}$ is the Fermi wavevector. Our experimentally determined $B_*$ requires $d\approx$25 nm. By contrast, the mean position of the edge resonances shows a relative insensitivity to $B$-field [Fig. \ref{Fig:Fig5}(c)], which could derive from the change in the shape (i.e. $d$ or $L$), or the enhanced local Fermi wavevector in the high-carrier density region at the edge. 
	
In conclusion, we have directly imaged resonant quasiparticle scattering in graphene. We have demonstrated that coherent scattering in disordered heterojunctions can be understood at a microscopic level by inspecting halo structures in scanning gate images. We have identified an important type of edge cavity effect induced by focusing of the electrostatic field. Our work both showcases the power of scanning probes at revealing the detailed behaviour of quantum devices and also paves the way towards imaging of coherence in ultra-high mobility and more sophisticated heterostructure devices based on two-dimensional atomic crystals.

We would like to acknowledge support from EPSRC and useful conversations with V. I. Fal'ko.

\renewcommand{\figurename}{Fig. S}

\section{Supplementary Information}

\subsection{Experimental Method}

Our graphene flakes are made by mechanically exfoliating natural graphite onto degenerately doped Si substrate with an oxide thickess of $\approx$300 nm. We identified monolayer flakes by their optical contrast and confirmed the thickness by measuring quantum Hall plateaus in a two-terminal configuration. Two- and three-terminal differential conductance measurements were taken using standard low frequency AC lock-in techniques, and a voltage $V_{BG}$ applied to the doped Si substrate controlled the carrier density. To enable charge writing we spin coated a 100 nm thick layer of PMMA over the device. The sample exhibited a Dirac point at V$_{BG}$ $\approx$ 20 V [red curve, Fig. 1(c)] and a carrier mobility of $\sim$3000 cm$^{2}$(Vs)$^{-1}$ at 2 $\times$ 10$^{11}$ cm$^{-2}$. Using the Einstein relation $\sigma=\nu e^2D$, where $\nu=8\pi \varepsilon_F/(h^2\nu_F^2)$ is the density of states at the Fermi level $\varepsilon_F\approx31$meV$\sqrt{V_{BG}}$ we find that $D\approx$0.03 m$^2$s$^{-1}$ is the diffusion constant, from which we estimate the electron mean-free path $l_{e}=2D/v_F$ to be approximately 80 nm. All measurements presented were carried out at a temperature of $\approx$4.2 K.

\subsection{Classical Transport and Electrostatic Model}

\begin{figure*}[!h]
\includegraphics[width=160mm]{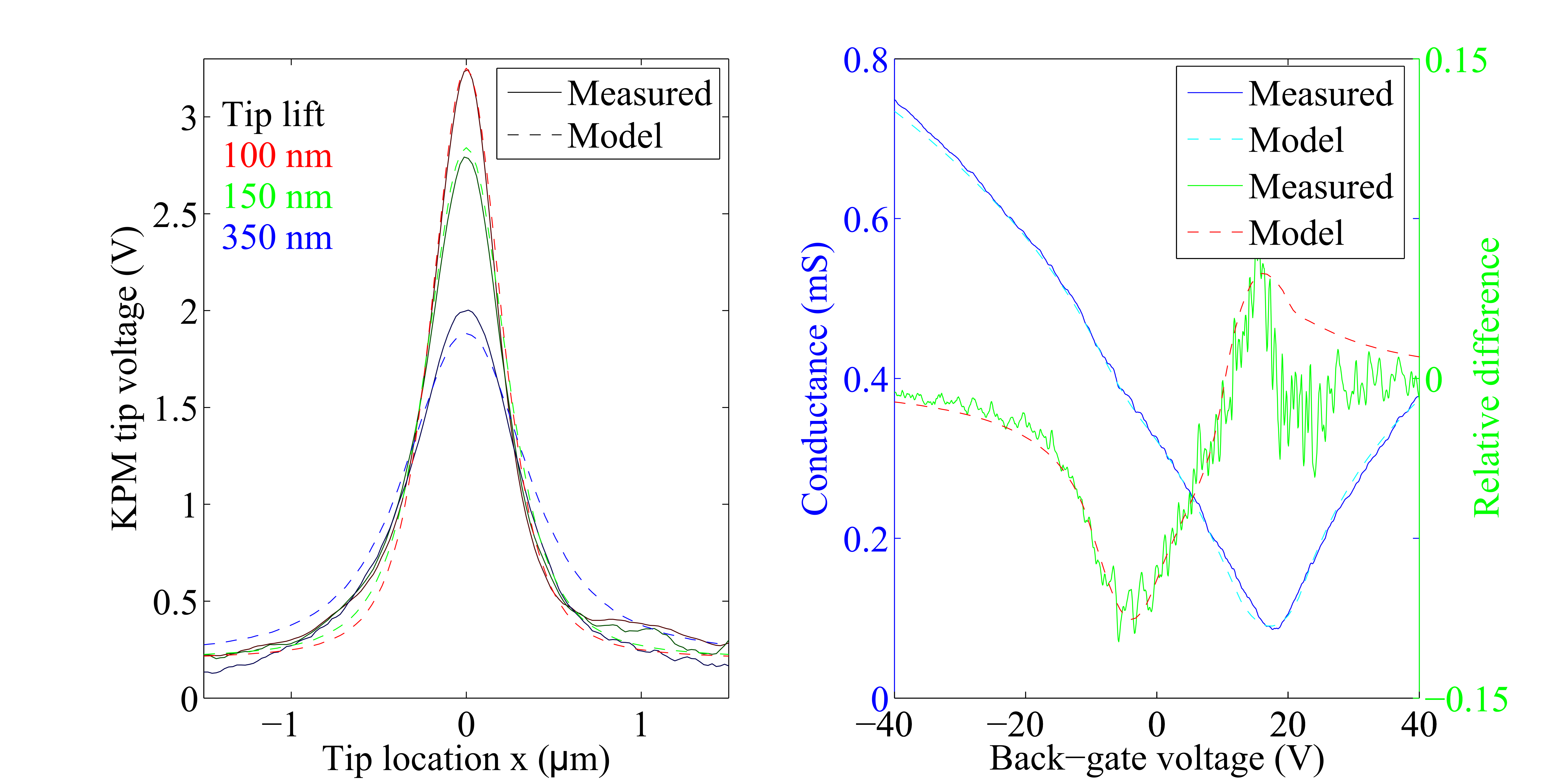}
\caption{(a) Cross sections of the surface potential measured using Kelvin probe microscopy for different tip heights over the PMMA surface. (b) Comparison between the measured and simulated conductance as a function of back-gate voltage before and after defining the heterojunction.}    
\label{Fig:FigS1}
\end{figure*}

To estimate the distribution of the deposited charge, the KPM and back-gate sweep measurements are simultaneously compared with their respective
models, and a search process is	used to iteratively approach the solution closest to the experimental results. The charge is first considered uniformly distributed in a rectangular volume, with the charge density and the dimensions of this volume left as free parameters. An error minimization process estimates the charge distribution related parameters. Firstly the length and the depth of the rectangle are initialised and the charge density is determined. Then, for the set of tip heights above the sample for which KPM measurements are available, the KPM line sections are simulated by solving the two-dimensional Laplace equation using Matlab. With the same initial parameters, the back-gate sweep, given the induced carrier density, is modelled using the method presented in Ref. \cite{Gorbachev2008a}. The discrepancy between the measured and simulated data define an error measure and minimizing this measure over a set of parameters leads to the best fits in Fig. S~\ref{Fig:FigS1}.

\subsection{Quantum Transport Model}

In order to numerically study the properties of the device for a
large range of potential landscapes, gate voltages and probe positions,
we have adopted a simplified simulation approach, replacing
a more exact, but time-consuming, self-consistent calculation with an
approximate calculation of the potential profile within the
device~\cite{marconciniiet2014}, which is then passed on to an
envelope-function based code for transport
simulation~\cite{logotetapre2014,marconcinijapdot2014}.

We start from the knowledge of the potential profile $U_0$
in the graphene layer for a particular set of voltages $V_{i_0}$ applied
to the gates~\cite{marconciniiet2014}. When the gate voltages are
changed by $\Delta V_i$, a variation $\Delta U$ results in the potential
profile (with respect to $U_0$), as well as a variation $\Delta \rho$ in the
charge density (with respect to the charge density $\rho_0$ corresponding
to the profile $U_0$). If the electrostatic coupling is modeled through
the capacitances $C_i$ (per unit area) between the gates and the flake, 
such variations are related by
\begin{equation}
\Delta \rho =\sum_i C_i \left( \frac{\Delta U}{-e}- \Delta V_i \right)
\label{electrostatics}
\end{equation}
(where $e$ is the modulus of the electron charge). On the other hand, the
charge density $\rho$ is directly related to the number of occupied states,
and thus to the local density of states. While the exact local density of
states depends on the wave function in the device and thus on the solution
of the transport problem, in the hypothesis of slow-varying potential it
can be approximated by shifting the argument of the density of
states by the local value of the potential energy. Under the 
further hypotheses of low temperature (Fermi-Dirac distribution approaching 
a step function), of quasi equilibrium (Fermi energy of the contacts nearly 
identical, equal to $E_F$), and of a sufficiently large graphene flake 
(density of states approaching that of unconfined graphene), the charge 
density can be expressed as
\begin{equation}
\rho=e \int_{E_F}^{U} {\rm DOS}(E-U)\,dE=
{\rm sign} (U-E_F) \frac{e}{\pi (\hbar v_F)^2}(U-E_F)^2\,,
\label{charge}
\end{equation}
with $U=U_0+\Delta U$ and $\rho=\rho_0+\Delta \rho$.  Substituting
Eq.~(\ref{electrostatics}) into Eq.~(\ref{charge}), a second-order equation
is obtained, which can be analytically solved in order to find the quantity
$\Delta U$ and thus the profile $U=U_0+\Delta U$.
Since in general the quantities $U$, $U_0$, $\Delta U$, $\rho$, $\rho_0$,
$\Delta \rho$, and $C_i$ are spatially-varying, the calculation has to be 
repeated for each point of the graphene flake. 

Then, the resulting approximate potential profile is passed on to the code for
transport simulation~\cite{logotetapre2014,marconcinijapdot2014}.
The structure is partitioned into a series of thin 
cascaded sections, in such a way that within each section the potential can 
be assumed as approximately constant along the transport direction. As a 
consequence, in each of these regions the envelope functions of graphene can 
be written as a confined transverse component multiplied by a longitudinally 
propagating plane wave. After some analytical manipulations, the resulting 
Dirac equation with Dirichlet boundary conditions is recast into a
differential equation with periodic boundary conditions, that can be
efficiently solved in the reciprocal space~\cite{fagottiprb2011,
marconcinijapsinc2014}. Then we enforce the continuity 
of the wave function at each interface between adjacent sections, on each of 
the two graphene sublattices and for all the possible modes impinging on the 
interface. Projecting these continuity equations on a basis of transverse 
functions, and solving the resulting linear system, the scattering matrix 
connecting the modes at the two sides of the interface is obtained. 
Recursively composing all the scattering matrices and applying the 
Landauer-B\"uttiker formula, we obtain the conductance of the overall structure.

With this approach, we have first simulated the transport behavior of the
graphene flake considering the effect of the back-gate, coupled
to the sample through a 0.1151~mF/(m$^2$) capacitance, and assuming a smooth
cavity-shaped potential $U_0$ with different profiles and widths.
Comparing the resistance behavior, and in particular the Fabry-P\'{e}rot
resonances resulting from the numerical simulations and from the experimental
measurements (see Fig.~2(b)), we have found a good agreement
assuming a Lorentzian profile $U_0$ with a 210~meV depth and a 180~nm width at
half maximum for $V_{BG}=0$~V.

\begin{figure*}
\centering
\includegraphics[width=\textwidth]{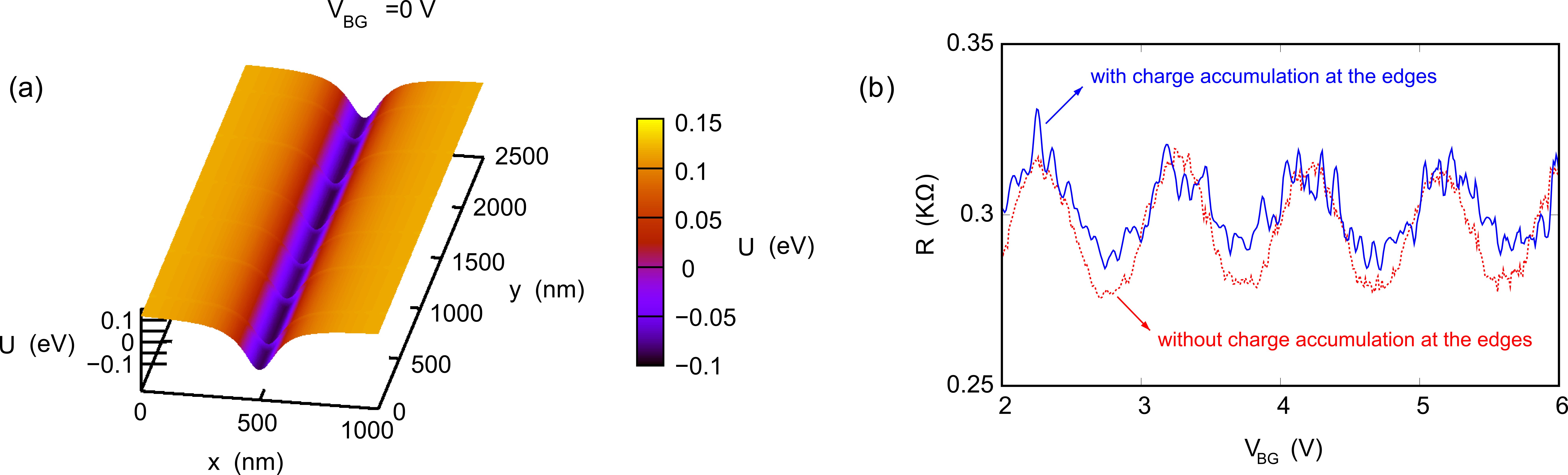}
\caption{(a) Potential profile, for $V_{BG}=0$~V, of the Lorentzian-shaped
cavity with longitudinal potential barriers considered in our simulations.
(b) Corresponding behavior of the resistance as a function of the
back-gate voltage, for a small range of $V_{BG}$ values.
}
\label{fig1}
\end{figure*}

Simulations have been performed also including 
the effect of potential disorder and other irregularities.
In particular, we have considered several longitudinal potential walls
with a height corresponding to about 10\% of the total depth of the cavity, 
that partition it into subcavities in parallel with an average
width of 400~nm (see the potential profile shown in Fig.~S~\ref{fig1}(a)). 
A finite dispersion is introduced in the values of the potential at the bottom
and of the width of the subcavities, as well as in the height of the walls separating
the cavities. The resulting behavior of the resistance as a function of the
back-gate voltage is reported with a dashed curve
in Fig.~S~\ref{fig1}(b), for a small range of $V_{BG}$ values.
Even though the disorder in the potential landscape introduces
irregularities in the resistance behavior, the Fabry-P\'{e}rot oscillations
typical of the original profile are still clearly visible.

We have then included the effect of electric field focusing at the edges,
obtained by solving Eq.~(\ref{charge}) with the complete Poisson equation for
a set of reference configurations and parametrizing the results as a function
of the back-gate voltage.
For the resulting potential profile, we have first repeated the
calculation of the resistance as a function of the back-gate voltage.
The results are shown with solid curves in Fig.~S~\ref{fig1}(b),
and are characterized by smaller and more rapid oscillations
(similar to those observed in the experiments), superimposed to the Fabry-P\'{e}rot
resonances already observed when the electric field focusing at the edges is
neglected.

Then, for this profile, we have performed a simulation of the 
resistance variation as the probe (located at a distance of 50~nm from 
the 100~nm thick dielectric coating and biased with a voltage $V_T=$-10~V) 
is scanned over the whole device. 
To this end, we have included into Eq.~(\ref{electrostatics}) a 
capacitance $C_T$ between the probe and each point $\vec r$ of the graphene 
flake, with a Lorentzian dependence on the  distance $d$ between $\vec r$ and 
the graphene point right underneath the probe: $C_T=C_{T_M}/(1+(d/d_0)^2)$, 
with $C_{T_M}=0.1171$~mF/(m$^2$) and $d_0=50$~nm.
The resistance values we have obtained are shown, as a function
of the probe position, in the lower panel of Fig.~2(c):
multiple halos appear, analogous to those observed in the experimental data
(see the upper panel of Fig.~2(c)).
These halos in the bulk of the flake originate from the formation of the
subcavities and, indeed, disappear if the scanning probe simulation is performed
on a single cavity (without the separation into several subcavities in
parallel).

We have then repeated the scanning probe simulation neglecting charge
accumulation at the edges of the flake. The difference between the resistance
values obtained with and without the effect of charge accumulation at the edges 
is shown, for a small subset of probe positions near the edge, 
in the upper right panel of Fig.~3(d).
From these results, it is apparent that electric field focusing 
leads to halos centered on the edges.

\end{document}